\documentclass[nato,namedreferences]{crckbked}    

\newcommand{\EQ}{\begin{equation}}
\newcommand{\EN}{\end{equation}}
\newcommand{\EQA}{\begin{eqnarray}}
\newcommand{\ENA}{\end{eqnarray}}
\newcommand{\eq}[1]{(\ref{#1})}
\newcommand{\EEq}[1]{Equation~(\ref{#1})}
\newcommand{\Eq}[1]{equation~(\ref{#1})}

\newcommand{\Fig}[1]{Fig.~\ref{#1}}

\newcommand{\Figss}[2]{Figs~\ref{#1}--\ref{#2}}

\newcommand{\bra}[1]{\langle #1\rangle}

\newcommand{\mean}[1]{\overline #1}
\newcommand{\meanemfs}{\overline{\cal E} {}}
\newcommand{\meanemf}{\overline{\mbox{\boldmath ${\cal E}$}} {}}
\newcommand{\meanAA}{\overline{\bf{A}}}
\newcommand{\meanBB}{\overline{\bf{B}}}
\newcommand{\meanJJ}{\overline{\bf{J}}}

\newcommand{\meanuu}{\overline{\mbox{\boldmath $u$}}{}}{}
{}
{}
{}
{}
{}
{}

\newcommand{\meanB}{\overline{B}}

\newcommand{\uu}{{\bf{u}}}
\newcommand{\BB}{{\bf{B}}}
\newcommand{\JJ}{{\bf{J}}}
\newcommand{\jj}{{\bf{j}}}
\newcommand{\AAA}{{\bf{A}}}
\newcommand{\aaaa}{{\bf{a}}}
\newcommand{\bb}{{\bf{b}}}

\newcommand{\kk}{{\bf{k}}}

\newcommand{\nab}{\mbox{\boldmath $\nabla$} {}}

\newcommand{\oo}{\mbox{\boldmath $\omega$} {}}

\newcommand{\dd}{{\rm d} {}}

\def\onethird{{\textstyle{1\over3}}}

\newcommand{\ycsf}[3]{: #1. {\em Chaos, Solitons \& Fractals} {\bf #2}, #3.}
\newcommand{\yepl}[3]{: #1. {\em Europhys. Lett.} {\bf #2}, #3.}

\newcommand{\yjgr}[3]{: #1. {\em JGR} {\bf #2}, #3.}

\newcommand{\yapj}[3]{: #1. {\em ApJ} {\bf #2}, #3.}

\newcommand{\yan}[3]{: #1. {\em AN} {\bf #2}, #3.}

\newcommand{\yana}[3]{: #1. {\em A\&A} {\bf #2}, #3.}

\newcommand{\ygafd}[3]{: #1. {\em Geophys. Astrophys. Fluid Dyn.} {\bf #2}, #3.}
\newcommand{\yjfm}[3]{: #1. {\em JFM} {\bf #2}, #3.}
\newcommand{\ypf}[3]{: #1. {\em Phys. Fluids} {\bf #2}, #3.}
\newcommand{\ypp}[3]{: #1. {\em Phys. Plasmas} {\bf #2}, #3.}
\newcommand{\ysov}[3]{: #1. {\em Sov. Astron.} {\bf #2}, #3.}
\newcommand{\yjetp}[3]{: #1. {\em Sov. Phys. JETP} {\bf #2}, #3.}

\newcommand{\yannr}[3]{: #1. {\em ARA\&A} {\bf #2}, #3.}

\newcommand{\yprl}[3]{: #1. {\em PRL} {\bf #2}, #3.}

\newcommand{\ymn}[3]{: #1. {\em MNRAS} {\bf #2}, #3.}
\newcommand{\ynat}[3]{: #1. {\em Nat} {\bf #2}, #3.}

\newcommand{\yjour}[4]{: #1. {\em #2}, {\bf #3.}, #4.}
\newcommand{\ybook}[3]{: #1. {\em #2} #3.}
\newcommand{\spr}[2]{: #1. {\em Phys. Rev.} {\bf #2}, submitted}
\newcommand{\sprl}[1]{: #1. {\em Phys. Rev. Lett.}, submitted}
\newcommand{\pjour}[2]{: #1. {#2}, in press}

\usepackage{graphicx,journals}

\begin{document}

\begin{article}

\begin{opening}

\title{MHD simulations of small and large scale dynamos}
\runningtitle{MHD simulations of small and large scale dynamos}

\author{A. \surname{Brandenburg}\email{brandenb@nordita.dk}}
\institute{Nordita, Blegdamsvej 17, DK-2100 Copenhagen \O, Denmark}

\author{N. E. L. \surname{Haugen}\email{nils.haugen@phys.ntnu.no}}
\institute{Dept.\ of Physics, The Norwegian University of Science
and Technology, H{\o}yskoleringen 5, N-7034 Trondheim, Norway}

\author{W. \surname{Dobler}\email{Wolfgang.Dobler@kis.uni-freiburg.de}}
\institute{Kiepenheuer-Institut f\"ur Sonnenphysik,
Sch\"oneckstra\ss{}e 6, D-79104 Freiburg, Germany}

\runningauthor{\surname{A. Brandenburg}, \surname{N. E. L. Haugen}
and \surname{W. Dobler}}


\begin{abstract}
Isotropic homogeneous hydromagnetic turbulence is studied using numerical
simulations at resolutions of up to $1024^3$ meshpoints.
It is argued that, in contrast to the kinematic regime, the nonlinear
regime is characterized by a spectral magnetic power that is
decreasing with increasing
wavenumber, regardless of whether or not the turbulence has helicity.
This means that the root-mean-square field strength converges to a
limiting value at large magnetic Reynolds numbers.
The total (magnetic and kinetic) energy spectrum tends to be somewhat
shallower than $k^{-5/3}$, in agreement with the findings of other groups.
In the presence of helicity, an inverse cascade develops, provided the
scale separation between the size of the computational box and the scale of the
energy carrying eddies exceeds a ratio of at least two.
Finally, the constraints imposed by magnetic helicity conservation on mean-field
theory are reviewed and new results of simulations are presented.
\end{abstract}

\end{opening}

\section{Introduction}

MHD turbulence is an important and ubiquitous phenomenon in astrophysics.
Stellar convection zones are known to harbor magnetic fields that can
be very strong locally, as evidenced by sunspots.
In the solar wind fully developed MHD turbulence has been measured in situ.
Accretion discs and galaxies are further examples where MHD turbulence
must be present.
Much of our current understanding of MHD turbulence has come from
analytical theories using a number of simplifications, and from
simulations at quite limited numerical resolution.
The time has now come to try and confirm various aspects
of MHD turbulence such as the power spectra of velocity and magnetic
field and the coupling between the two via dynamo action
in numerical simulations at higher resolution, with $1024^3$ meshpoints
being feasible now.

One important question is how intermittent the magnetic field is
and whether it is dominated by large or small scale structures.
This question is directly related to observable quantities in the
interstellar medium where the contributions from large and small scale
field are found to be comparable; see Beck et al.\ (1996) for a review.
The question of small scale fields is also related to theoretical
approaches to understand the amplification of the large scale magnetic
fields on scales ranging from stars to galaxies in general, as well as the
cyclic modulation of such fields on late type stars such as the sun.

\section{Isotropic MHD turbulence}

We are particularly interested in the case where the magnetic
field is not externally maintained, but constantly regenerated by
dynamo action.
Dynamo action is nothing special, but a rather general phenomenon that
is common to virtually all turbulent and also many non-turbulent flows.
One important requirement is that the magnetic Reynolds number is large,
i.e.\ that resistive effects are weak and that the field is nearly
frozen into the flow at large and moderate scales.
Only under rather special conditions can dynamo action be suppressed,
so for example if the magnetic field is forced to be two-dimensional.

\subsection{The kinematic growth phase}

There are normally always two phases to dynamo action: the kinematic
phase where the field is still weak and the dynamical phase where
the velocity is being affected by the Lorentz force which then
leads to the saturation of the magnetic field.
It is worth recalling that there can be exceptions to this rather
simple rule: there are self-killing and self-generating dynamos
(Fuchs et al.\ 1999).
The self-generating dynamos are relevant to accretion discs where the
turbulence driving the dynamo is only generated by the magnetic field
itself via the magneto-rotational instability (Balbus and Hawley 1991).
This requires a doubly positive feedback from both the magneto-rotational
and the dynamo instability.
Simulations of this phenomenon (Brandenburg et al.\ 1995) have not only
confirmed that this process works, but they have also provided one of
the clearest examples of large scale dynamo action in general, and
migratory cyclic behavior in particular.
The geodynamo is another example where a finite amplitude magnetic
field is essential for the operation of the dynamo on the so-called
strong-field branch that is thought to be relevant for the Earth
(Glatzmaier and Roberts 1995).

\begin{figure}[t!]
\centering
\includegraphics[width=.99\textwidth]{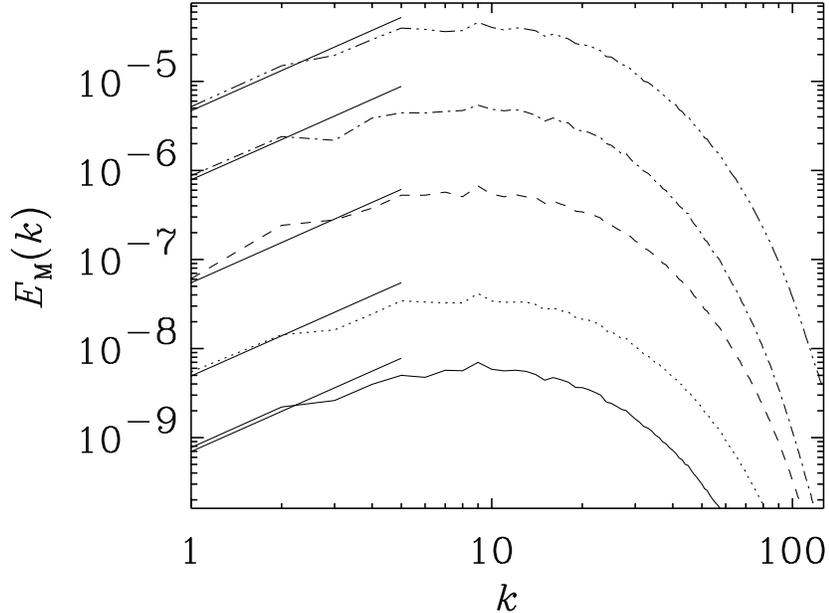}
\caption{Magnetic power spectra during the early growth phase of a forced
  MHD turbulence simulation.
The straight lines denote the $k^{3/2}$ Kazantsev slope.}
\label{plot_evol_powerb2_early}
\end{figure}

In the following we return to the more standard situation where the
turbulence exists already in the absence of magnetic fields.
In this section we restrict ourselves to the case where the turbulence is
non-helical, so there is no well established mechanism
that could generate large scale fields
(the $\alpha$-effect, which is discussed in the following sections,
is absent).
The field grows then predominantly at small scales.
This process is well described by the theory of Kazantsev (1968)
which predicts that the magnetic energy spectrum, $E_{\rm M}(k)$, increases
with $k$ like
\EQ
E_{\rm M}(k)\sim k^{3/2}\quad\mbox{(kinematic regime)}.
\EN
In simulations of forced MHD turbulence this behavior has been demonstrated in
the case of large magnetic Prandtl numbers (Maron and Cowley 2001).
In this case the kinetic energy cascade terminates much before the
magnetic energy cascade.
Recent simulations of Haugen et al.\ (2003) have shown that the same
behavior is also found in the case of unit magnetic Prandtl number,
$\mbox{Pr}_{\rm M}\equiv\nu/\eta=1$; see \Fig{plot_evol_powerb2_early}.

\subsection{The saturated phase}

\begin{figure}[t!]
\centering
\includegraphics[width=.99\textwidth]{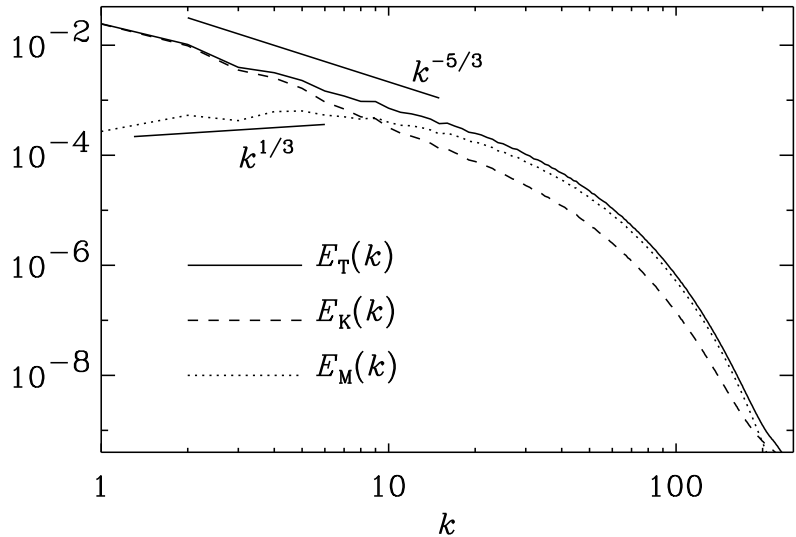}
\caption{Kinetic and magnetic energy spectra compared
with the total energy spectrum
(dashed, dotted, and solid lines, respectively) for the saturated state of
a forced MHD turbulence simulation.
At small wavenumbers, the spectral magnetic energy increases
approximately like $k^{1/3}$ (at higher resolution, however,
the spectrum is flatter; see Haugen et al.\ 2003).
The forcing is nonhelical with amplitude $f_0=0.02$,
$\nu=\eta=2\times10^{-4}$ and $512^3$ meshpoints.}
\label{FET_EM_EK_512}
\end{figure}

\begin{figure}[t!]
\centering
\includegraphics[width=.99\textwidth]{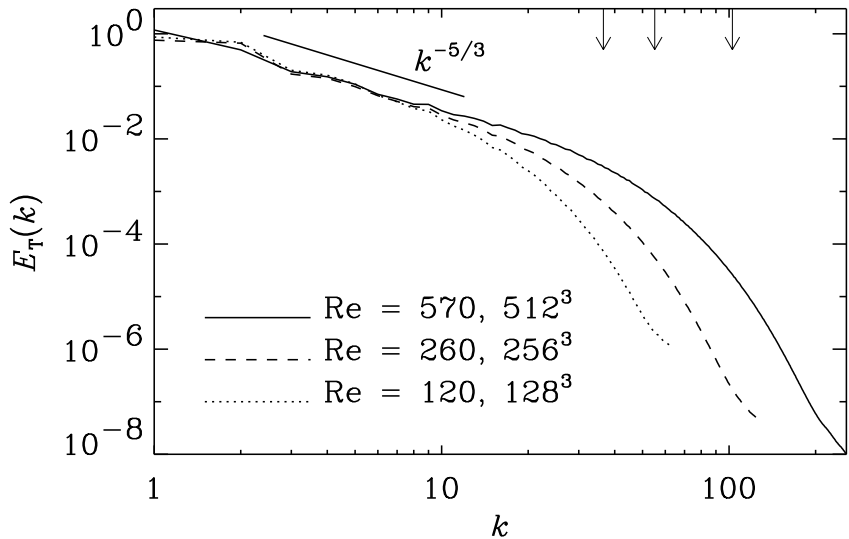}
\caption{Comparison of total energy spectra for runs
with different Reynolds numbers. Stronger forcing: $f_0=0.05$.
The vertical arrows at the top give the visco-resistive cutoff
wavenumber, $(\epsilon/\nu^3)^{1/4}$.
}
\label{Fkolmogorov_nocomp}
\end{figure}

\begin{figure}[t!]
\centering
\includegraphics[width=.99\textwidth]{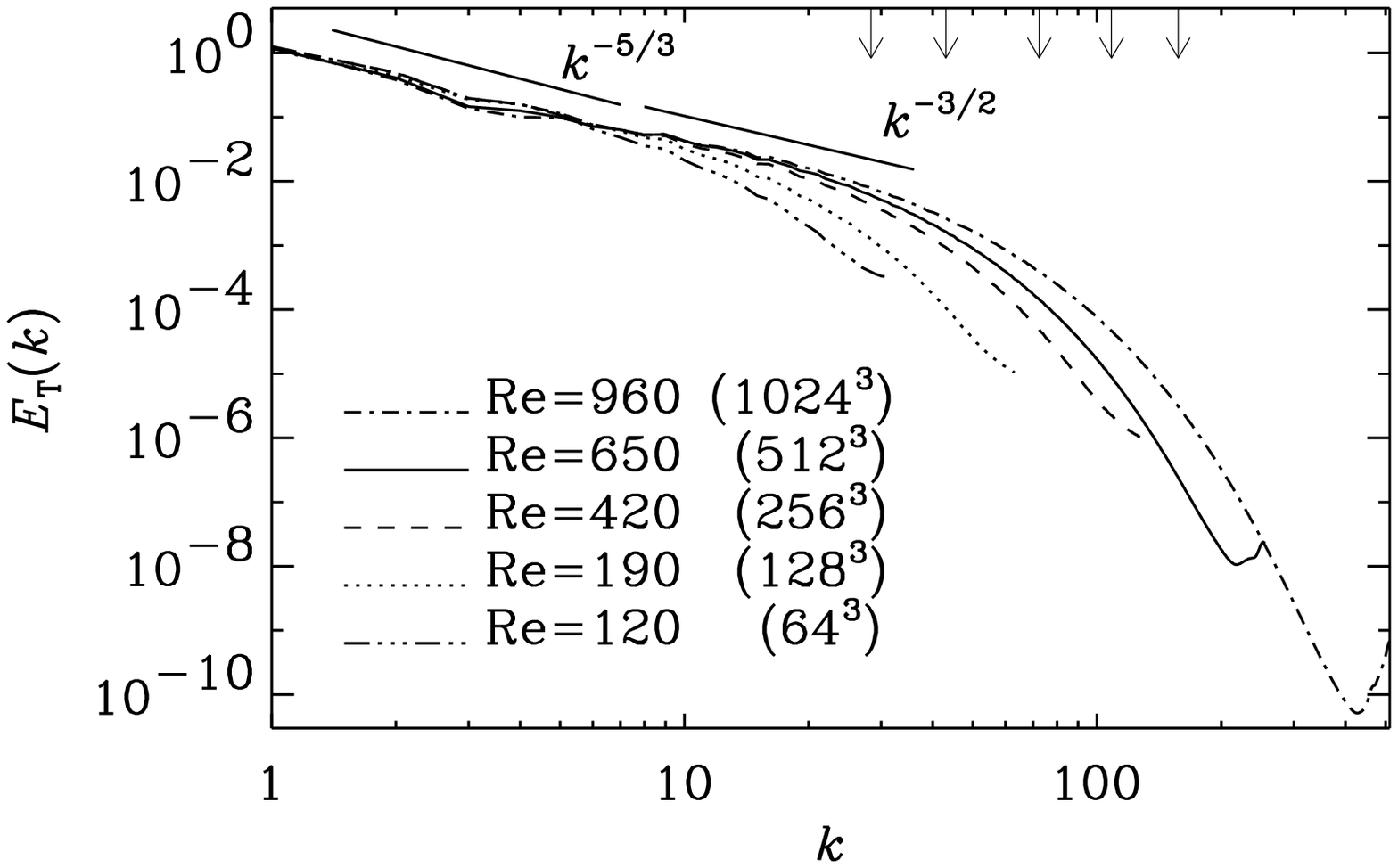}
\caption{Comparison of total energy spectra for runs
with different Reynolds numbers. Weaker forcing: $f_0=0.02$.
The `hook' to the right of the high resolution spectrum
indicates that the run has not yet sufficiently relaxed yet
and is typical of similar runs shortly after remeshing
from a lower resolution run.
}\label{Fkolmogorov_nocomp2}
\end{figure}

The rapid build-up of small scale fields, as seen in linear (kinematic)
theory, has often been viewed as an indication that also in the
nonlinear, saturated phase the field is dominated by small scale fields
(e.g.\ Kulsrud and Andersen 1992); see Widrow (2002) for a review.
This picture might be in conflict with the ordinary picture
of a turbulent cascade where an approximately scale-independent
energy flux $\epsilon$ gives rise to energy transfer from large to small scales
(e.g.\ Frisch 1995).
This question has indeed been matter of recent debate.
For decades the leading theory of MHD turbulence has been that
of Iroshnikov (1963) and Kraichnan (1965).
They suggested that in the inertial range the
kinetic and magnetic power spectra are
given by
\EQ
E_{\rm M}(k)=E_{\rm K}(k)=C_{\rm IK}\,\epsilon^{1/2} v_{\rm A}^{1/2}k^{-3/2}
\quad\mbox{(nonlinear regime)},
\EN
where $v_{\rm A}$ is the Alfv\'en speed and $C_{\rm IK}$ is a constant.
Until recently this spectrum seemed to be verified by simulations
(Biskamp 1993, 1995, Biskamp and Bremer 1994), but solar wind data
always suggested a $k^{-5/3}$ spectrum (Matthaeus and Goldstein 1982)
as in Kolmogorov theory.
This issue has been resolved by Goldreich and Sridhar (1995, 1997)
who showed that the magnetic field makes the flow strongly anisotropic
(especially on small scales) and that an energy cascade occurs only for
the components perpendicular to the field lines.
This picture was also supported by simulations with imposed fields
(Maron and Goldreich 2001, Cho and Vishniac 2000a, Cho et al.\ 2002a;
see also Oughton et al.\ 1998)
and with decaying fields (Biskamp and M\"uller 2000).

The existence of the  Goldreich--Sridhar cascade requires the presence
of reasonably strong large scale fields, and it is not clear whether
dynamo-generated fields satisfy this condition.
Simulations of nonhelical dynamos by Maron and Cowley (2001) and
Cho and Vishniac (2000b) do not seem
to show power law behavior for the magnetic power spectrum, but rather a
peak in the spectrum near the resistive scale that would
move to larger wave numbers (smaller
scales) as the magnetic Reynolds number is increased.

A different suggestion was made by Kida et al.\ (1991) who
found from a dynamo simulation that the {\it total} energy spectrum,
$E_{\rm T}(k)\equiv E_{\rm M}(k)+E_{\rm K}(k)$, is of the form
\EQ
E_{\rm T}(k)=C_{\rm KYM}\epsilon^{2/3}k^{-5/3},
\EN
where $\epsilon=\epsilon_{\rm M}+\epsilon_{\rm K}$ is the sum of magnetic
and kinetic energy dissipation rates, and $C_{\rm KYM}\approx2.1$ is the
`Kolmogorov' constant found by Kida et al.\ (1991).

Simulations by Haugen et al.\ (2003) at resolutions of up to $1024^3$
meshpoints seem to confirm that the magnetic energy spectrum does not
show power law behavior, but only the kinetic energy spectrum and
the total energy spectrum.
The precise form of the power spectrum is not fully settled however;
cf.\ \Figss{FET_EM_EK_512}{Fkolmogorov_nocomp2} for results with
different forcing amplitudes and Reynolds numbers.

Already at a resolution of $512^3$ (\Fig{Fkolmogorov_nocomp}) it
becomes evident that the $k^{-5/3}$ spectrum is followed by a shallower
spectrum just before entering the dissipation range.
At a resolution of $1024^3$ meshpoints (\Fig{Fkolmogorov_nocomp2}),
the emergence of a shallower spectrum ($k^{-3/2}$ or perhaps even
$k^{-1}$) becomes even more pronounced.
This result is very similar to the ``bottleneck effect'' seen quite
strongly in simulations using hyperviscosity and hyperresistivity
(Cho and Vishniac 2000b, Biskamp and M\"uller 2000).
Here the ordinary $\nabla^2$ operator is replaced by an operator
$(-1)^{n-1}\nabla^{2n}$, where $n$ is an integer usually
taken between 2 and 8.
The bottleneck effect exists also in two-dimensional turbulence even
with ordinary viscosity and resistivity (Biskamp et al.\ 1998).
It is worthwhile recalling that a $k^{-1}$ bottleneck subrange
is also observed in simulations where the magnetic
Prandtl number is large (Cho et al.\ 2002b).

The emergence of a shallower $k^{-1}$ subrange has also been
found in hydrodynamic simulations with numerical dissipation
via the piecewise parabolic method (PPM); see Porter et al.\ (1992).
Again, this subrange has become ever more pronounced
as the resolution was increased to $1024^3$ meshpoints
(Porter et al.\ 1998).
PPM is believed to act similarly to hyperviscosity, so the emergence of
a $k^{-1}$ bottleneck subrange seems therefore not too surprising.
In the present simulations with magnetic fields, however, this explanation
does not apply, because ordinary viscosity and magnetic
diffusion operators are used.
The nature of a shallower $k^{-3/2}$ or even $k^{-1}$ subrange in our
MHD simulations is therefore unclear.
One wonders what will happen at much larger Reynolds numbers.
Will the spectrum then be dominated by a long $k^{-1}$ subrange,
or is this just an artifact of still too small Reynolds numbers?
Or is this a subtle feature of meshpoint schemes being run too closely
to the maximum permissible value of the Reynolds number?
On the other hand, the bottleneck effect may well be physical, and the
reason why this possibility is emerging only now is that it is much
weaker in the one-dimensional longitudinal or transversal spectra
accessible from experiments as compared to the fully three-dimensional spectra
available in simulations (Dobler et al.\ 2003).
In the following section we attempt a preliminary convergence study by
comparing runs at different numerical resolution and different values
of the Reynolds number.

\section{Reynolds number dependence and convergence}

\begin{figure}[t!]
\centering
\includegraphics[width=.99\textwidth]{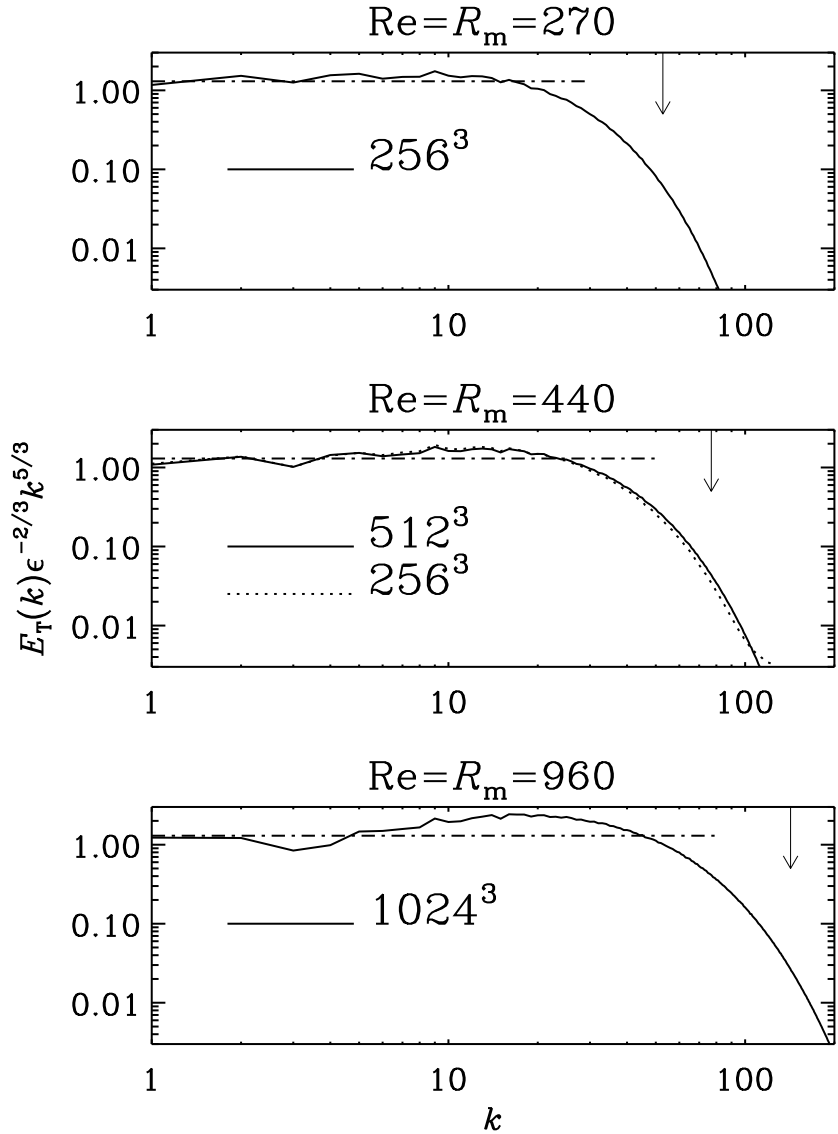}
\caption{Comparison of compensated total energy spectra for runs
with different Reynolds numbers.
In all runs the horizontal dash-dotted line goes through
the value 1.3.
In the second panel two runs with the same Reynolds numbers,
but different resolution are compared.
$f_0=0.02$.
}\label{conv_test}
\end{figure}

In turbulence research it is customary to display compensated power
spectra, $E(k)/(\epsilon^{2/3}k^{-5/3})$.
If the spectrum obeys the anticipated scaling one expects the
compensated spectrum to be flat over a certain range.
The value of the spectrum gives the Kolmogorov constant or, in the
present case, $C_{\rm KYM}$, because, following Kida et al.\ (1991),
we are considering here the total (kinetic plus magnetic) energy
spectrum and use the rate of total energy dissipation,
$\epsilon=\epsilon_{\rm M}+\epsilon_{\rm K}$.
The result is shown in \Fig{conv_test} for three different values
of $\mbox{Re}=R_{\rm m}$ between 270 and 960.
In the first case with $\mbox{Re}=R_{\rm m}=270$ the compensated
spectrum shows a reasonably flat range with $C_{\rm KYM}\approx1.3$,
which is less than the value found by Kida et al.\ (1991).
In the second case with $\mbox{Re}=R_{\rm m}=440$ the compensated
spectrum shows clear indications of excess power just before turning
into the dissipative subrange.
This effect becomes even more dramatic in the third case with
$\mbox{Re}=R_{\rm m}=960$.

In order to assess the reliability of the results we have carried
out a convergence study using the same value of $\nu$, but different
mesh resolution; see the second panel of \Fig{conv_test}.
In both cases, $\nu=\eta=2\times10^{-4}$, while $u_{\rm rms}=0.13$
for $512^3$ meshpoints and 0.12 for $256^3$.
The energy dissipation is also similar, $\epsilon=2.8\times10^{-4}$
for $512^3$ meshpoints and $2.3\times10^{-4}$ for $256^3$.
The compensated energy spectra agree quite well for the two different
resolutions, and both show excess power just before turning into the
dissipative subrange.
This supports the result that the excess power may be real.

\section{When helical and nonhelical turbulence are similar}

\begin{figure}[t!]
\centering
\includegraphics[width=.99\textwidth]{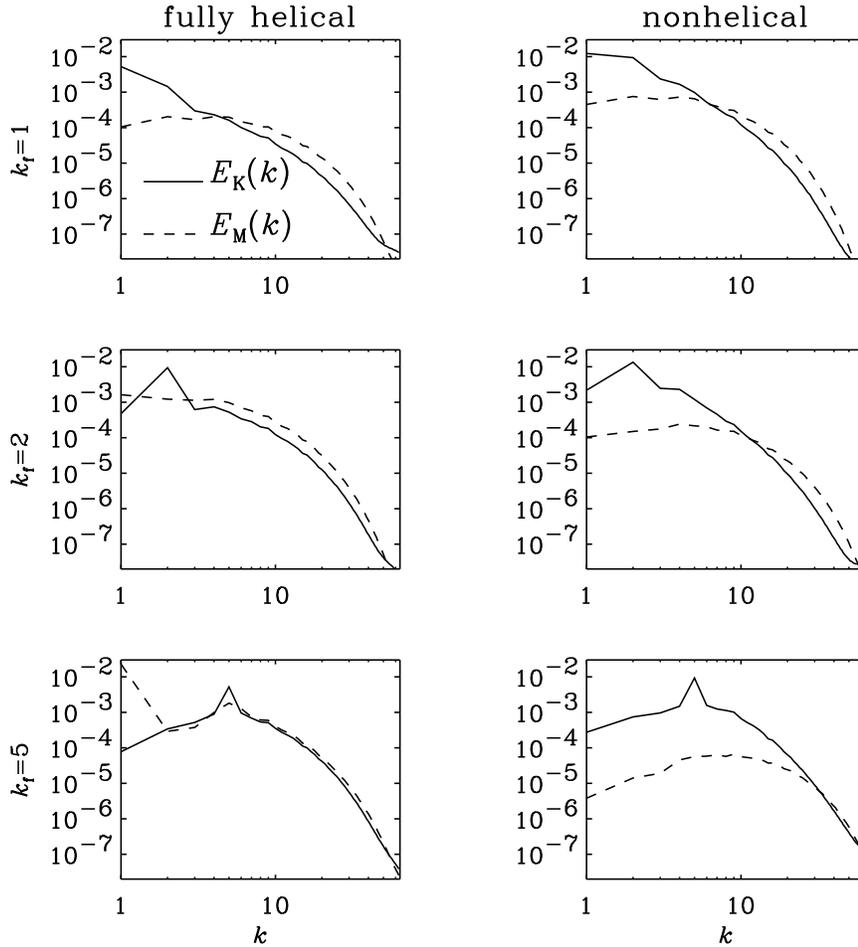}
\caption{Comparison of runs with and without helicity
(left and right hand columns, respectively) and for
the different forcing wavenumbers:
$k_{\rm f}=1$ (top) $2$ (middle), $5$ (bottom).}
\label{budapest1}
\end{figure}

It is important to point out that 
the simulations of Kida et al.\ (1991) are not strictly
comparable to our present case, because they considered
the case of helical forcing.
In the following we shall discuss
the difference between helical and nonhelical turbulence.

In the case of helical forcing one expects an {\it inverse} cascade to
smaller wavenumbers rather than a direct cascade to larger wavenumbers.
This is not really seen in the simulations of Kida et al.\ (1991).
There are two reasons for this.
On the one hand the inverse cascade takes a resistive
time to develop (Brandenburg 2001, hereafter B01),
and this time tends to be too long
if magnetic hyperdiffusivity is used (Brandenburg and Sarson 2002).
Kida et al.\ (1991) did use magnetic hyperdiffusivity and their resistive
times where at least a hundred times longer than the duration of their
runs, so no inverse cascade should be expected.
But there is another, perhaps more important reason.
In order for the inverse cascade to develop, scale
separation is required, i.e.\ the magnetic field must be allowed to grow on
scales larger than the forcing scale
(which corresponds to the energy carrying scale) of
the turbulence.
This was not the case in the simulations of
Kida et al.\ (1991), and thus, there is no inverse cascade present in
their simulations.
Therefore their results should be close to those without helicity.
This is shown more clearly in \Fig{budapest1} where we compare the
power spectra of helical and nonhelical simulations with forcing at
wavenumber $k_{\rm f}=1.5$ (no scale separation), $k_{\rm f}=2.3$
(weak scale separation), and $k_{\rm f}=5$ (considerable scale separation).
For $k_{\rm f}=1.5$ the spectra of the helical and nonhelical simulations
are indeed quite similar to each other.

\section{Inverse cascade and nonlinear $\alpha$-effect}

The magnetic field generated by the inverse cascade tends to have helicity
of opposite signs at large and small scales.
This is a consequence of magnetic helicity conservation which, in the
limit of large magnetic Reynolds numbers, prevents
the build-up of any net magnetic helicity.
Forcing with positive helicity, for example, results
in positive magnetic helicity at small scales (in particular the forcing scale).
To conserve magnetic helicity, this positive magnetic helicity
must be compensated by negative magnetic helicity at large scales, which
in turn is a
direct consequence of the inverse cascade, through which magnetic energy flows
from small to large scales.

Another way of seeing this is by noting that helical turbulence leads
to an $\alpha$-effect.
This means that the large scale magnetic field, $\meanBB$, can be
described by the mean-field dynamo equation (Moffatt 1978)
\EQ
\partial\meanBB/\partial t=
\nab\times(\alpha\meanBB-\eta_{\rm T}\nab\times\meanBB),
\label{dynamo_eqn1}
\EN
where $\eta_{\rm T}=\eta_{\rm t}+\eta$ is the total (sum of turbulent
and microscopic) magnetic diffusivity.
In periodic domains, this equation has solutions where
$\alpha\meanBB=\eta_{\rm T}\nab\times\meanBB$, so
\EQ
\mu_0\mean\JJ\cdot\meanBB=(\alpha/\eta_{\rm T})\meanBB^2,
\EN
where $\mean\JJ=\nab\times\meanBB/\mu_0$ is the mean current.
Thus, for these solutions the sign of the current helicity density of the
large scale field, $\mean\JJ\cdot\meanBB$, is the same as the sign
of $\alpha$.

The problem with mean-field theory is that not only the magnitude
(and sometimes even the sign) of $\alpha$ are not well understood,
but in particular that the feedback from the magnetic field has
completely ignored the effects of the small scale fields
(Vainshtein and Rosner 1991, Vainshtein and Cattaneo 1992).

\subsection{The helicity constraint}

\begin{figure}[t!]
\centering
\includegraphics[width=.99\textwidth]{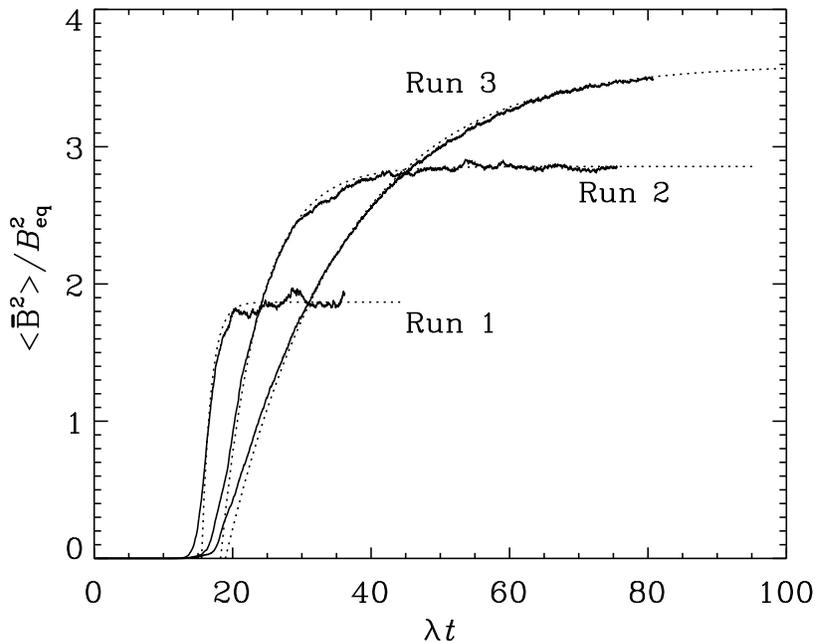}
\caption{Late saturation phase of fully helical turbulent dynamos
for three different values of the magnetic Reynolds number:
$R_{\rm m}\equiv u_{\rm rms}/\eta k_{\rm f}=2.4$, 6, and 18 for
Runs~1, 2, and 3 respectively; see B01.
The mean magnetic field, $\meanBB$, is normalized with
respect to the equipartition value,
$B_{\rm eq}=\sqrt{\mu_0\rho_0}u_{\rm rms}$,
and time is normalized with respect to the kinematic
growth rate, $\lambda$.
The dotted lines represent the fit formula \eq{helconstraint}
which tracks the simulation results rather well.}
\label{pjbm_decay_nfit_budap}
\end{figure}

In this section we briefly explain how the question of feedback
is directly related to magnetic helicity conservation;
see Eq.~(\ref{magn_hel_eqn}) below.
Even without thinking about mean-field theory, the magnetic helicity
balance poses a problem when it comes to building up a large scale
field fast.
In simulations with helicity, the build-up of large scale magnetic
field is always seen to proceed on a resistive timescale.
Ideally, one would like to see such built-up to proceed on a dynamical
timescale, but so far there have been no simulations that are actually
able to generate large scale fields faster than on the resistive timescale.
A resistively slow saturation phase means that the resistivity, or
magnetic diffusivity $\eta$, enters the timescale of the saturation.
This is demonstrated in \Fig{pjbm_decay_nfit_budap} where we show the
saturation phase of a turbulent dynamo together with the fit formula
\EQ
{\meanBB^2\over B_{\rm eq}^2}
={\epsilon_{\rm f}k_{\rm f}\over\epsilon_{\rm 1}k_{\rm 1}}
\left(1-e^{-2\eta k_1^2(t-t_{\rm sat})}\right)
\label{helconstraint}
\EN
that was derived in B01, see also Brandenburg et al.\ (2002)
using the assumptions that
the large and small scale fields are nearly fully helical, i.e.\
that the helicity fractions $\epsilon_1$ and $\epsilon_{\rm f}$ of the
mean and fluctuating fields, respectively, are nearly 100\%.

The resistively slow saturation behavior, which is also seen in
oscillatory dynamos with shear ($\alpha\Omega$-type dynamos; see
Brandenburg et al.\ 2001)
is not reproduced by standard mean-field dynamo theory.
Only in the special case of an $\alpha^2$-type dynamo (no shear)
it has been possible to describe the saturation behavior correctly
by simultaneously invoking $R_{\rm m}$-dependent $\alpha$ and
$\eta_{\rm t}$ quenchings (B01).
For an extensive review including a discussion of the effects of shear,
helicity cancellation across the equator, and the effect on the
cycle frequency, see Brandenburg et al.\ (2002).
Here we just want to emphasize that magnetic helicity conservation
is a real issue and not just a theoretical complication.

\subsection{Small-scale magnetic helicity evolution}

The next question is how to deal with magnetic helicity conservation
and how to incorporate it into mean-field models.
The answer is relatively simple if the system is homogeneous and there
are no boundaries, i.e.\ if the boundary conditions are periodic.
In that case we have to satisfy
\EQ
\dd\bra{\AAA\cdot\BB}/\dd t=-2\eta\mu_0\bra{\JJ\cdot\BB}.
\label{magn_hel_eqn}
\EN
Here, $\AAA$ is the magnetic vector potential, and $\bra{\AAA\cdot\BB}$
is the magnetic helicity, with angular brackets denoting volume
averages.\footnote{In a periodic domain $\bra{\AAA\cdot\BB}$ is gauge
invariant, because $\bra{\nab\phi\cdot\BB}=\bra{\phi\nab\cdot\BB}=0$}
The mean-field dynamo equation \eq{dynamo_eqn1}
may also be written in the form
\EQ
\partial\meanBB/\partial t=
\nab\times(\meanemf+\meanuu\times\meanBB-\eta\mu_0\meanJJ),
\label{dynamo_eqn2}
\EN
where $\meanemf=\alpha\meanBB-\eta_{\rm t}\mu_0\meanJJ$ is the mean
electromotive force expressed in its simplest form in terms of the mean
magnetic field and its curl via $\alpha$-effect and turbulent diffusion.
Mean fields are here defined as averages over one or at most two
coordinate directions.
In the case of the $\alpha^2$ dynamo the wavevector of the mean field,
$\kk_{\rm m}$, may point in any of the three coordinate directions,
so the averages should be taken over the two directions perpendicular
to $\kk_{\rm m}$.

\EEq{dynamo_eqn2} implies its own evolution equation for the magnetic
helicity of the mean field, $\bra{\meanAA\cdot\meanBB}$,
\EQ
\dd\bra{\meanAA\cdot\meanBB}/\dd t=2\bra{\meanemf\cdot\meanBB}
-2\eta\mu_0\bra{\meanJJ\cdot\meanBB}.
\EN
Note that any field-aligned component of $\meanemf$ produces {\it excess}
magnetic helicity at large scales relative to the magnetic helicity of
the total field.
This can only be consistent with \Eq{magn_hel_eqn} if magnetic helicity
of opposite sign is absorbed by the small scale (or fluctuating) field.
We shall now quantify this.

The departures from the mean field are referred to as fluctuations, i.e.\
$\bb=\BB-\meanBB$ and $\aaaa=\AAA-\meanAA$.
This implies
\EQ
\bra{\aaaa\cdot\bb}=\bra{\AAA\cdot\BB}-\bra{\meanAA\cdot\meanBB}
\EN
and
\EQ
\bra{\jj\cdot\bb}=\bra{\JJ\cdot\BB}-\bra{\meanJJ\cdot\meanBB}
\EN
and hence the evolution equation for the magnetic helicity of
the fluctuations is
\EQ
\dd\bra{\aaaa\cdot\bb}/\dd t=-2\bra{\meanemf\cdot\meanBB}
-2\eta\mu_0\bra{\jj\cdot\bb}.
\label{dabdt}
\EN
This equation plays a key role in the dynamical $\alpha$-quenching
approach that was first developed by Kleeorin and Ruzmaikin (1982)
and turns now out to be the only theoretically acceptable theory that
reproduces the helicity constraint seen in simulations (see Field and
Blackman 2002, Blackman and Brandenburg 2002, Subramanian 2002).
We shall show now that
this equation can be coupled back into the mean-field equation
by noting that the quantities $\bra{\aaaa\cdot\bb}$ and $\bra{\jj\cdot\bb}$
are proportional to magnetic contributions to the $\alpha$-effect.

\subsection{Calculating the $\alpha$-effect}

A new way of deriving the expression coupling the mean electromotive
force with the mean field has recently been put forward by
Blackman and Field (2002) who calculated the time derivative of $\meanemf$,
\EQ
{\dd\meanemfs_i\over\dd t}=\epsilon_{ijk}\left(\overline{u_j\dot{b}_k}
+\overline{\dot{u}_jb_k}\right),
\label{demfdt}
\EN
where
\EQ
\dot{\bb}=\nab\times(\uu\times\meanBB)+...,
\quad\nab\cdot\bb=0,
\EN
\EQ
\dot{\uu}=-\nab p+\bb\cdot\nab\meanBB+\meanBB\cdot\nab\bb+...,
\quad\nab\cdot\uu=0,
\EN
and diffusive and nonlinear terms have been ignored
(first order smoothing approximation).
Inserting this into \Eq{demfdt} yields
\EQ
{\dd\meanemfs_i\over\dd t}=\tilde\alpha_{ij}\meanB_j
-\tilde\beta_{ijk}\meanB_{j,k}-{\meanemfs\over\tau},
\label{demfdt}
\EN
where all terms beyond the (linear) first order smoothing
approach have been subsumed into the last term, which acts
as a damping term provided $\tau$ is positive.
The positivity is indeed confirmed using simulations
(Brandenburg and Blackman, unpublished).
The calculation of $\tilde\alpha_{ij}$ is relatively
straightforward and yields (cf.\ Blackman and Field 2002)
\EQ
\tilde\alpha_{ip}=\epsilon_{jnp}\overline{u_j u_{n,i}}
-\epsilon_{inp}\overline{u_j u_{n,j}}
+\epsilon_{ijk}\overline{b_kb_{j,p}}.
\EN
If the assumption of isotropy is made one has
$\tilde\alpha_{ip}=\tilde\alpha\delta_{ij}$ with
\EQ
\tilde\alpha=
-\onethird\overline{\oo\cdot\uu}
+\onethird\overline{\jj\cdot\bb}/\rho_0.
\label{alpha}
\EN
One usually ignores the explicit time derivative of $\meanemf$ in \Eq{demfdt},
in which case $\alpha=\tau\tilde\alpha$ is the well-known $\alpha$-effect
(e.g.\ Krause and R\"adler 1980).
The magnetic contribution, $\tau\,\overline{\jj\cdot\bb}/(3\rho_0)$,
was first derived by Pouquet et al.\ 1976, and used heavily in all
recent approaches to $\alpha$-quenching (Gruzinov and Diamond 1994,
Bhattacharjee and Yuan 1995, Field et al.\ 1999, Blackman and Field 2000).

\subsection{The dynamical quenching model}

\begin{figure}[t!]
\centering
\includegraphics[width=.99\textwidth]{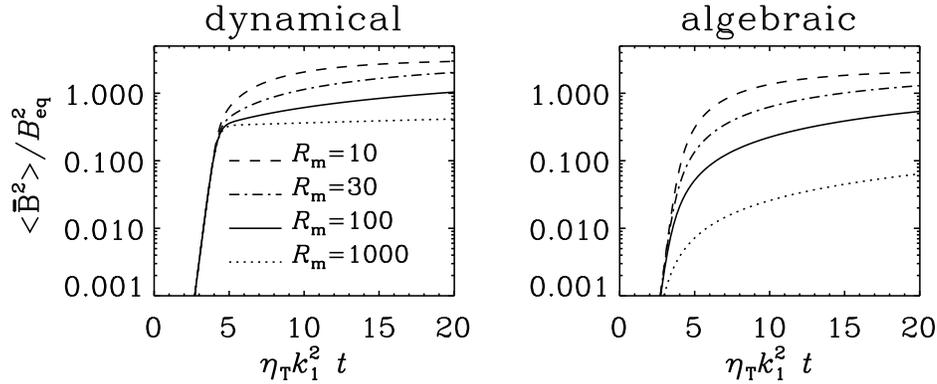}
\caption{Comparison of the magnetic field evolution in the dynamical
and algebraic quenching models for different values of the magnetic
Reynolds number.
Both models have in common that $\eta_{\rm t}$ is assumed to be
proportional to $\alpha$ at all times.
While both models agree in their late saturation behavior, the onset of
saturation is markedly more abrupt in the dynamical quenching model.
}\label{pcomp2}
\end{figure}

We have seen that the evolution equation for the small scale
magnetic helicity involves the term $\bra{\jj\cdot\bb}$, which
is also part of the $\alpha$-effect; see \Eq{alpha}.
Here we have also made use of the assumption of isotropy.
This assumption, together with the assumption of scale separation,
allows us to write
\EQ
\bra{\aaaa\cdot\bb}\approx\bra{\jj\cdot\bb}/k_{\rm f}^2.
\EN
This equation can be used to write \Eq{dabdt} fully in terms of $\alpha$,
\EQ
{\dd\alpha\over\dd t}=-2\eta k_{\rm f}^2
\left(R_{\rm m}{\bra{\meanemf\cdot\meanBB}\over B_{\rm eq}^2}
+\alpha-\alpha_{\rm K}\right),
\label{dynquench}
\EN
where $\alpha_{\rm K}=\onethird\tau\bra{\oo\cdot\uu}$ is the kinematic
contribution to the $\alpha$-effect.
\EEq{dynquench} has first been obtained by Kleeorin and Ruzmaikin
(1982), and it has been generalized by Kleeorin et al.\ (2000, 2002)
to incorporate the flux of small scale magnetic helicity through the
boundaries.
In the special context of homogeneous systems this model was rederived and
analyzed by Field and Blackman (2002) and Blackman and Brandenburg (2002).

The most important property of \Eq{dynquench} is that it reproduces the
late, resistively slow saturation phase in a physically motivated fashion.
It is for this reason that an explicitly time-dependent $\alpha$-effect
appears to be mandatory for all models of large scale dynamos based
on the helicity (or $\alpha$) effect.

\subsection{Catastrophic quenching as a limiting case}

The time-dependent equations \eq{demfdt} and \eq{dynquench}
for $\alpha$ and $\meanemf$ may seem rather unfamiliar.
In order to make contact with earlier more familiar formulations we now
ignore the explicitly time dependence in \Eq{dynquench} and obtain
\EQ
\alpha=\alpha_{\rm K}-R_{\rm m}{\bra{\meanemf\cdot\meanBB}\over B_{\rm eq}^2}
\equiv\alpha_{\rm K}-\alpha R_{\rm m}{\bra{\meanBB^2}\over B_{\rm eq}^2}
+\eta_{\rm t}R_{\rm m}{\bra{\meanJJ\cdot\meanBB}\over B_{\rm eq}^2},
\EN
where $\meanemf=\alpha\meanBB-\eta_{\rm t}\meanJJ$ itself depends still
on $\alpha$.
Solving for $\alpha$ yields
\EQ
\alpha={\alpha_{\rm K}
+\eta_{\rm t}R_{\rm m}\bra{\meanJJ\cdot\meanBB}/B_{\rm eq}^2
\over1+R_{\rm m}\bra{\meanBB^2}/B_{\rm eq}^2}.
\EN
This equation was already obtained by Gruzinov and Diamond (1994)
and Kleeorin et al.\ (1995).
In numerical experiments with an imposed large scale field
over the scale of the box, $\meanBB$ is spatially uniform and
hence $\meanJJ=0$.
In that special case we have
\EQ
\alpha={\alpha_{\rm K}\over1+R_{\rm m}\bra{\meanBB^2}/B_{\rm eq}^2}
\quad\mbox{($\meanJJ=0$)},
\EN
which implies that $\alpha$ becomes quenched when
$\bra{\meanBB^2}/B_{\rm eq}^2=R_{\rm m}^{-1}\approx10^{-8}$
for the sun.
This would seem to preclude any involvement of $\alpha$ in the sun
(Vainshtein and Cattaneo 1992).
However, if the field is not imposed but maintained by dynamo action,
$\meanBB$ is not spatially uniform and then $\meanJJ$ is finite. 
In the case of a Beltrami field (see B01),
$\bra{\meanJJ\cdot\meanBB}/\bra{\meanBB^2}=\tilde{k}_{\rm m}$
is some effective wavenumber of the large scale field.
Since $R_{\rm m}$ enters both the nominator and the denominator,
$\alpha$ tends to $\eta_{\rm t}\tilde{k}_{\rm m}$.
Thus, the question of how strongly $\alpha$ is quenched in the sun
has been diverted to the question of how strongly $\eta_{\rm t}$
is quenched (Blackman and Brandenburg 2002).
One way to determine this quantity is by looking at cyclic dynamos
with shear ($\alpha\Omega$-type dynamos), because here the cycle
frequency is equal to $\eta_{\rm t}\tilde{k}_{\rm m}^2$.
Blackman and Brandenburg (2002) came to the conclusion that $\eta_{\rm t}$
is quenched when $\bra{\meanBB^2}/B_{\rm eq}^2$ becomes of order unity,
i.e.\ it is noncatastrophically quenched.
However, more work at larger magnetic Reynolds numbers needs to be done.

\subsection{Comparison with simulations}

At magnetic Reynolds numbers up to a few tens the comparison between
simulations and the dynamical quenching model turned out to be quite
favorable.
However, as we increase the magnetic Reynolds number to one hundred and
above a new difficulty arises that was already visible in Run~5 of B01.
The point is that both during the entire kinematic phase and during much of
the saturation phase several modes of the large scale field with the
same or very similar values of $|\kk|$ are excited and contribute to
the evolution of the mean field.
Which one survives in the end can remain undecided for quite a long time when
the magnetic Reynolds number is large.
This is also the main reason why in \Fig{pjbm_decay_nfit_budap} the runs
with smaller magnetic Reynolds number reached equipartition earlier than
those with larger magnetic Reynolds number.
This can be seen from Fig.~6 of B01 which shows the interchange of energy
in different modes during the intermediate and late saturation phase.

\begin{figure}[t!]
\centering
\includegraphics[width=.99\textwidth]{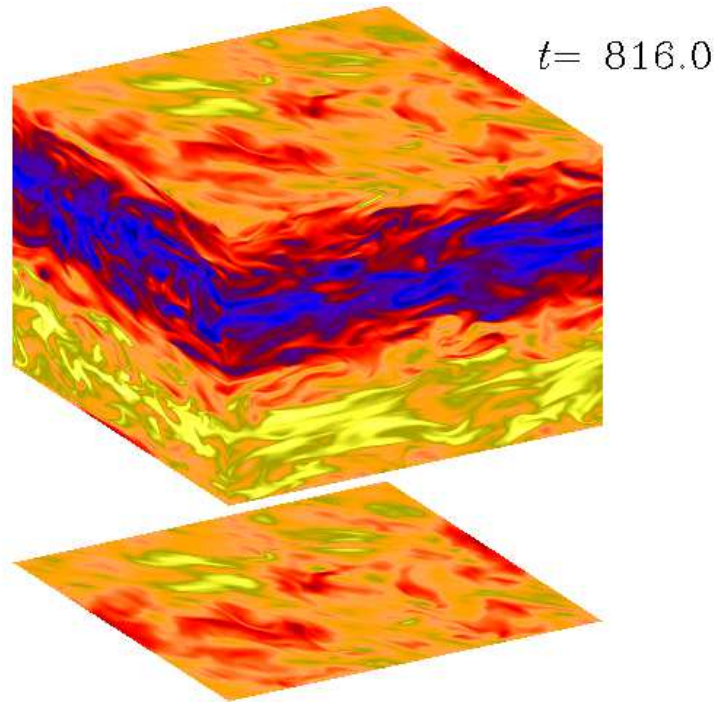}
\caption{Snapshot of $B_x$ from a run with helical forcing,
showing the modulation of $B_x$ in the $z$ direction.
Note the elongation of structures in the directions
perpendicular to the $z$ direction.
The initial field is a Beltrami field of low amplitude.
$512^3$ meshpoints.
}\label{img_0010}
\end{figure}

\begin{figure}[t!]
\centering
\includegraphics[width=.99\textwidth]{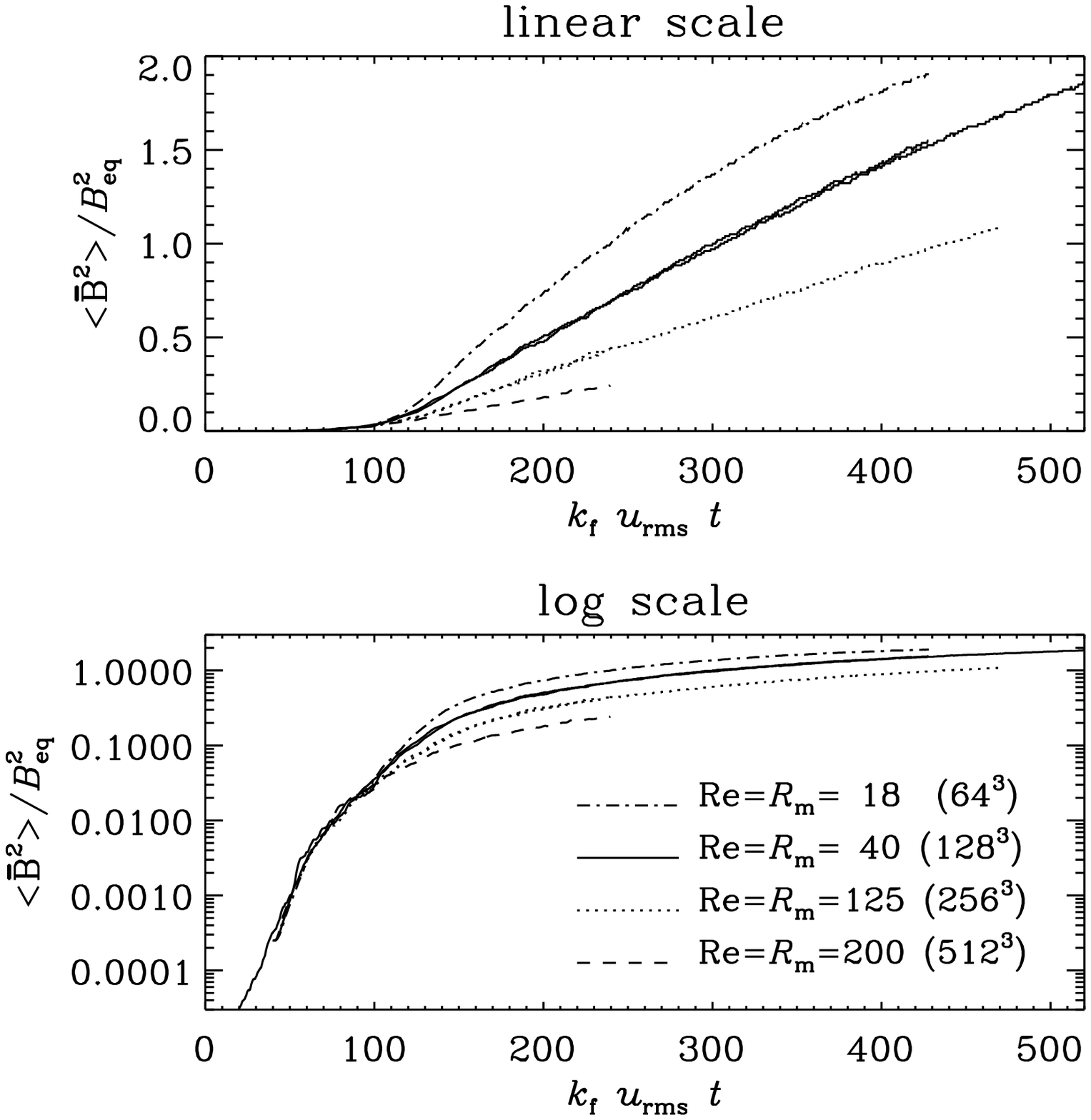}
\caption{Evolution of the mean magnetic field in simulations
with different magnetic Reynolds numbers starting from an initial Beltrami
field.
Runs for the same $R_{\rm m}$, but with different initial field
strengths result in overlapping curves.
In these runs $u_{\rm rms}\approx0.2$ and $k_{\rm f}=5$.
}\label{pn_comp_shift}
\end{figure}

One can artificially try to favor a particular mode by initializing the
simulation with a strong Beltrami field; see \Fig{img_0010}.
This was done in a series of runs shown in \Fig{pn_comp_shift}, where we
show the evolution of the mean field for magnetic Reynolds numbers up
to about $R_{\rm m}\equiv u_{\rm rms}/(\eta k_{\rm f})=200$.
(Here, $k_{\rm f}=5$, so $R_{\rm m}$ is, in this definition, $\sim3$ times
smaller than with $k_{\rm f}=1.5$ and the same resolution.)
Although the late saturation behavior agrees fairly well with the
dynamical quenching model,
at intermediate times the agreement is less favorable.
An obvious possibility that needs to be considered is
$R_{\rm m}$-independent (non-catastrophic) quenching of $\alpha$ and
$\eta_{\rm t}$.
In Blackman and Brandenburg (2002) such quenching for $\eta_{\rm t}$
was already found to be vital for explaining simulations with shear.

\subsection{Comments on models without helicity}

Given the severe constraints encountered in connection with the
helicity effect, one has to ask the question whether or not astrophysical
large scale dynamos do operate with the helicity effect, or whether
there are other effects that produce large scale fields without
producing net magnetic helicity.
Several alternative effects have been mentioned in recent years.
Vishniac and Cho (2001) suggested a new effect that might operate in
accretion discs where shear is very strong.
Simulations by Arlt and Brandenburg (2001) have so far not been
able to show that such an effect would be powerful enough to
generate large scale fields.
A similar effect is R\"adler's (1969) $\Omega\times\meanJJ$ effect
(see also Krause and R\"adler 1980),
and Kitchatinov (these proceedings) has given reasons why this effect
may indeed be important in the sun.
Finally, we mention negative magnetic diffusivity effects
(Zheligovsky et al.\ 2001), but again it is not clear that
this actually works in astrophysical turbulence.
We emphasize, however, that these are possibilities that should not
be overlooked.

\section{Concluding remarks}

The present investigations have demonstrated that a lot can be learned
from numerical simulations at different resolutions going up to the highest
resolution possible today, which is around $1024^3$ meshpoints.
An important question is whether these simulations have all run for
long enough to give reliable statistics.
The spectra vary a lot in time and averaging the spectra over long enough
time intervals is crucial.
In addition, there is the worry that the entire spectrum may evolve
slowly in time after remeshing from a lower resolution run.
It seems, however that at least the simulations with $512^3$ meshpoints,
which have generally run for between 50 to 80 (large-scale) turnover
times have now reached a statistically steady state.
The simulation with $1024^3$ meshpoints, however, has run for only 5
turnover times, but even that may be long enough, although one can not
be certain.
We also recall that, as we have shown in this paper, simulations at
different resolution and with different initial conditions give the
same result.

Another question is what can be learned from simulations at intermediate
Reynolds numbers when one is really interested in asymptotically large
Reynolds numbers like those for the sun and many other astrophysical bodies.
It is possible that the shallower subrange just before the dissipative
subrange would disappear as the Reynolds number becomes larger.
This possibility would be reminiscent of what happened in the mid-eighties,
when hydrodynamic simulations began to show that turbulence
is permeated by many thin and long vortex tubes
(Kerr 1985, She et al.\ 1990, Vincent and Meneguzzi 1991).
These tubes have a diameter comparable to the dissipative scale and their
length was thought to be equal to the integral scale, which in turn is
comparable to the size of the computational domain.
Only recently, simulations at resolutions up to $1024^3$ meshpoints
(Porter et al.\ 1998) have begun to show that the length of the tubes
is significantly less than the integral scale.
Instead, visualizations of vorticity show that clusters of vorticity
are now a more dominant feature of high Reynolds number turbulence.
Thus, there may well be surprises ahead of us in our quest for a deeper
understanding of astrophysical MHD turbulence.

\begin{acknowledgements}
{\footnotesize
Use of the supercomputers in Odense (Horseshoe), Trondheim (Gridur), 
and Leicester (Ukaff) is acknowledged.
}
\end{acknowledgements}


\bibliographystyle{klunamedv}  
\bibliography{solar}           


\end{article}
\end{document}